\documentclass[12pt,preprint]{aastex}

\shorttitle{Signature of the Ice Line and Modest Migration}
\shortauthors{Schlaufman et al.}

\begin{document}

\title{The Signature of the Ice Line and Modest Type I Migration in the
Observed Exoplanet Mass-Semimajor Axis Distribution}

\author{Kevin C. Schlaufman\altaffilmark{1} and D. N. C. Lin\altaffilmark{2}}
\affil{Astronomy and Astrophysics Department, University of California,
Santa Cruz, CA 95064; kcs@ucolick.org and lin@ucolick.org}

\and

\author{S. Ida}
\affil{Tokyo Institute of Technology, Ookayama, Meguro-ku, Tokyo 152-8551,
Japan; ida@geo.titech.ac.jp}

\altaffiltext{1}{NSF Graduate Research Fellow}
\altaffiltext{2}{Kavli Institute of Astronomy and Astrophysics, Peking
University, Beijing, China}

\begin{abstract}
Existing exoplanet radial velocity surveys are complete in the planetary
mass-semimajor axis ($M_p-a$) plane over the range 0.1 AU $< a <$ 2.0 AU where
$M_p \gtrsim 100~M_{\oplus}$.  We marginalize over mass in this complete
domain of parameter space and demonstrate that the observed $a$ distribution
is inconsistent with models of planet formation that use the full Type I
migration rate derived from a linear theory and that do not include the effect
of the ice line on the disk surface density profile.  However, the efficiency of
Type I migration can be suppressed by both nonlinear feedback and the barriers
introduced by local maxima in the disk pressure distribution, and we confirm
that the synthesized $M_p-a$ distribution is compatible with the observed data
if we account for both retention of protoplanetary embryos near the ice line
and an order-of-magnitude reduction in the efficiency of Type I migration. The
validity of these assumption can be checked because they also predict a
population of short-period rocky planets with a range of masses
comparable to that of the Earth as well as a ``desert" in the $M_p-a$
distribution centered around $M_p \sim 30-50~M_{\oplus}$ and $a <1$ AU.
We show that the expected ``desert" in the $M_p-a$ plane  will be
discernible by a radial velocity survey with 1 m s$^{-1}$ precision and
$n \sim 700$ radial velocity observations of program stars.
\end{abstract}


\section{Introduction}

Over 200 planets with reliable mass ($M_p$) and semimajor axis ($a$)
measurements have been discovered around nearby FGK stars in the past decade.
At the same time, attempts to build a comprehensive deterministic
theory of planet formation have lead to the development of population
synthesis models based on the sequential accretion scenario.
In \citet{ida04}, two of us studied the growth of planetesimals
into dynamically isolated embryos as well as their tidal interactions with their
parent disks.  Using the observed ranges of disk mass, size, and accretion
rate we showed that a fraction of embryos evolve into cores with more than a
few Earth masses ($M_\oplus$), accrete massive envelopes, open up gaps near
their orbits, and attain asymptotic masses comparable to that of Jupiter.
In some massive and persistent disks, the newly formed gas giant planets may
migrate toward the proximity of their host stars.  In the end these
simulations produced the distribution of dynamical and structural properties of
planets.  Presently, the observed sample of extrasolar planetary properties
has become large enough to enable direct comparisons between the theoretically
predicted and observed $M_p-a$ distributions that not only delineate the 
dominant physical mechanisms at work in planet formation, but also
provide quantitative constraints on the efficiencies of those processes.

In the latest update of the planet formation models two of us have
incorporated the effect of Type I migration \citep{ida08a}. This
process is a direct consequence of a protoplanetary core's tidal
interaction with its parent protoplanetary disk.  The efficiency of Type I
migration was first determined by a linear theory \citep{gol80,war86,tan02}
that neglected the embryo's perturbation on the
surface density distribution of its parent protoplanetary disk.
In the environment of a minimum mass nebula though, this efficiency factor
would imply that a protoplanetary embryo with mass a fraction of an Earth mass
would migrate from $\sim 1$ AU into its host star
before the severe depletion of the disk gas that is known to occur over a time
scale of several Myr.  Although this critical mass at which Type I migration
causes AU-scale migration on the disk depletion time increases with $a$,
it is still difficult to retain sufficiently massive cores for the onset of
dynamical accretion of gas.  This argument implies that gas giants should
be very rare \citep{ida08a} and this paradox has led to many in-depth
analyses of the Type I migration process.  Numerical nonlinear simulations of
Type I migration were reviewed by \citet{pap06} and many potential
explanations for slow Type I migration are in the literature: intrinsic
turbulence in the disk \citep{lau04,nel04}, self-induced unstable flow
\citep{kol04,li05}, nonlinear radiative and hydrodynamic feedback
\citep{mas06a}, and variation in surface density and temperature gradients
\citep{mas06b}.  \citet{dob07} has shown that in some
situations the nonlinear Type I migration rate can be less than 10\%
of the linear prediction.

\defcitealias{ida08b}{IL} 	
Another issue studied by two of us \citep[IL hereafter]{ida08b} is the
critical embryo mass $M_{crit} >$ at least a few $M_{\oplus}$ required by
current models for runaway gas accretion.  The embryo is limited by its
isolation mass $M_{iso}$, and the solid surface density profile of the minimum
mass solar nebula (MMSN) $\Sigma_d \propto a^{-3/2}$ requires that the
$M_{iso}$ scales like $a^{3/4}$.  On the other hand, the timescale for growth
$\tau_{c,acc}$ scales with $a^{27/10}$.  As a result, after the
characteristic gas depletion time $\tau_{dep}$ the most massive
embryos near the ice line have masses $M_c \sim M_{iso} < M_{crit}$.
However, since we observe many exoplanets with Jupiter masses at $a \sim 1$
AU, there must be some physical process that is neglected in this
simple analysis.  \citet{kre07} outlined one possible solution to this
problem in which solids are trapped near regions of the disk where the
local pressure requires the gas to rotate with super-Keplarian
velocities.  When the combined contribution of both Lindblad and
corotation resonances are taken into account \citep{mas06b}, the
migration of protoplanetary cores may be suppressed as well
\citepalias{ida08b}.

In this paper, we utilize the observed data to calibrate the
population synthesis models.  In \S2 we quantitatively show that the
existing distribution of exoplanets cannot be explained by models of
planet formation that apply the full Type I migration rate predicted
from linear theory and that do not include the effects of the ice
line.  We also point out that the existing observed and synthesized $M_p-a$
distributions are in agreement with each other if we take into account
the effect of an ice line barrier and assume a reduction in the magnitude
of Type I migration.  In addition, we describe the parameters of a radial
velocity survey capable of verifying the existence of ``desert" in the $M_p-a$
diagram predicted by \citetalias{ida08b}.  In \S3 we consider the implications
of these models and suggest methods to test our assumptions.  In \S4 we
summarize our findings.

\section{Analysis}

We use the ideas presented in \citet{nar05} and the formalism developed in
\citet{cum04} to approximately reproduce the result of \cite{cum08} that
showed that the current California and Carnegie Planet Search
(CCPS) has announced all planets with orbital period $P < 2000$ days, stellar
reflex velocities $K > 20$ m s$^{-1}$, and eccentricities $e \lesssim 0.6$.  In
particular, we determine which of the simulated exoplanet systems from
\citetalias{ida08b} would be detectable by a radial velocity survey with
precision and cadence similar to present-day radial velocity surveys like
the CCPS and the High Accuracy Radial Velocity Planetary Search Project
(HARPS).  In this approach, we combine Equation (26) and (30) of \cite{cum04}
and declare that all the synthesized planetary systems from \citetalias{ida08b}
with mass $M_p > M_{50}$ where

\begin{eqnarray}\label{eq1}
M_{50} & \approx & \frac{70~M_{\oplus}}{\sqrt{N}\sin{i}}
                   \left(\frac{\sigma}{\mbox{m/s}}\right)
                   \left(\frac{P}{\mbox{yr}}\right)^{1/3}
                   \left(\frac{M_{\ast}}{M_{\odot}}\right)^{2/3}
                   \left[\frac{\ln{(M/F)}}{9.2}\right]^{1/2}
\end{eqnarray}

\noindent
are detectable.  In Equation~(\ref{eq1}), $N$ is the number
of radial velocity observations, $\sigma$ is the RMS of spectrograph
precision and stellar jitter, $i$ is the inclination
of the exoplanet's orbit, $P$ is the orbital period in years, and $M_{\ast}$
is the host stellar mass in solar masses.  In the limit of large $N$,
$M \approx 100$ is the number of independent frequencies searched and
$F \approx 0.01$ is the false alarm probability -- the numerical values are
correct to order-of-magnitude and in any case they only very weakly influence
our estimate of $M_{50}$.  We note that Equation~(\ref{eq1}) is formally
correct only for single planetary systems in circular orbits; however,
\citet{cum04} shows that in the limit of large $N$ Equation~(\ref{eq1}) applies
to multiple planet systems and in the case where $e \lesssim 0.6$.  Therefore,
we set $M_{50} = \infty$ if $e > 0.6$.  We assume a velocity resolution of
1 m s$^{-1}$.

We then carry out a Monte Carlo simulation in which we assign each of the
simulated planetary systems from \citetalias{ida08b} a random host stellar mass,
stellar jitter, eccentricity, inclination, and number of radial velocity
observations.  In this prescription, we use the empirical distributions for
host stellar mass, stellar jitter, eccentricity, and number of radial
velocity observations given in the updated \citet{but06}
catalog\footnote{Maintained at http://www.exoplanets.org} of all known
exoplanets; we use the standard
distribution for random inclinations. We compute which planetary systems have
$M_p > M_{50}$ and declare that these systems are detectable in this iteration.

\begin{figure}
\plotone{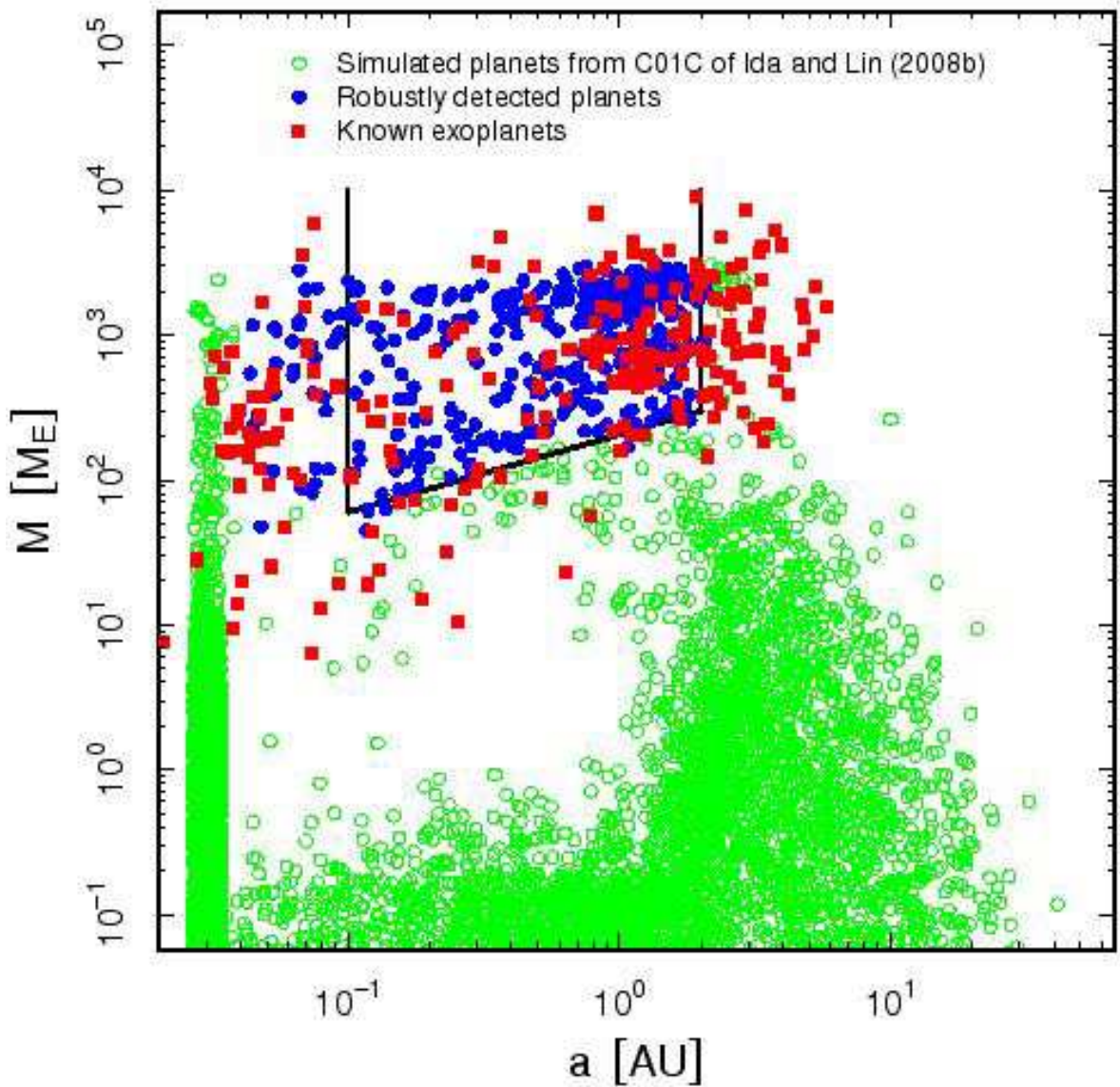}
\caption{Results of our Monte Carlo simulation.  The open circles are
the simulated planets from Figure 3 of \citet{ida08b}; the model includes
the effects of the snow line on the surface density of gas $\Sigma_g$ and
dust $\Sigma_d$.  The Type I migration rate in the simulation is 0.1 times
the prediction from linear theory.  The filled circles are planets that are
robustly detected, that is, planets that would be detected at least 90\% of
the time by current radial velocity surveys.  We also plot all known exoplanet
planets as filled squares.  All simulated planets in the range
0.1 AU $< a <$ 2.0 AU with $M_p$ as specified by Equation~(\ref{eq2}) denoted
by the heavy black lines are robustly detected, so existing radial velocity
surveys are complete in that range.  See the electronic edition of the Journal
for a color version of this figure.\label{fig1}}
\end{figure}

We repeat this process 1000 times.  We consider all planetary systems
that are detectable according to the $M_{50}$ criterion in 90\% of the
Monte Carlo iterations robustly detectable.  Averaged over host
stellar mass, stellar jitter, eccentricity, inclination, and number of
radial velocity observations we find that all of the simulated planets
from \citetalias{ida08b} in the range 0.1 AU $< a <$ 2.0 AU with

\begin{eqnarray}\label{eq2}
M_p & \gtrsim & \left[174\left(\frac{a}{\mbox{au}}\right)+47\right] M_{\oplus}
\end{eqnarray}

\noindent
are robustly detectable.  We include the results of this calculation in
Figure~\ref{fig1}.  The robustly detectable systems are marked with solid
circles and we plot the observed $M_p-a$ distribution of known planetary
systems as the solid squares.

\begin{figure}
\plotone{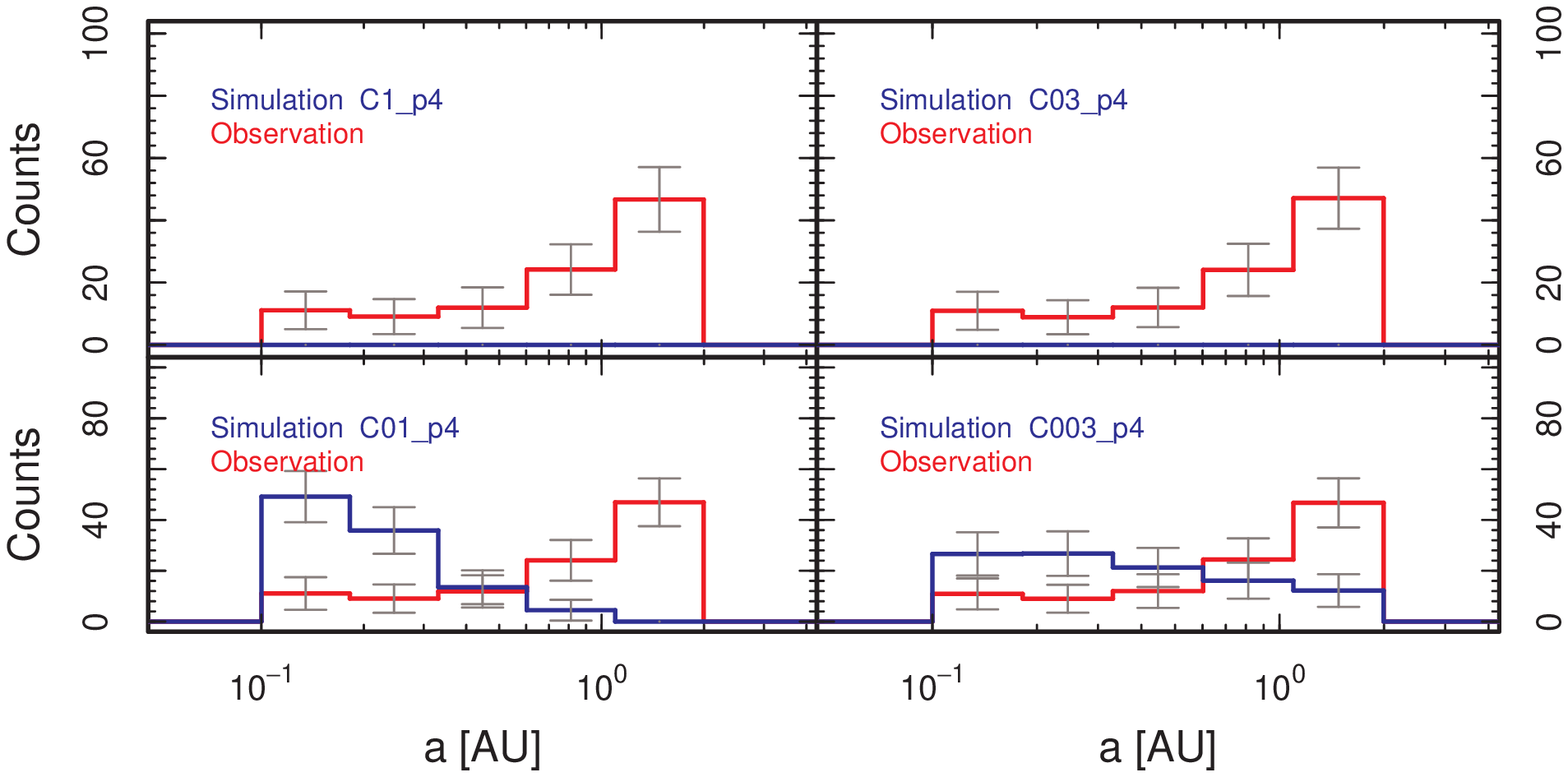}
\caption{1D distribution in semimajor axis derived from the projection
of the 2D distribution in the complete region of the $M_p-a$ plane from
\citet{ida08b} for the disks without the bump in $\Sigma_g$ or $\Sigma_d$.
The upper left panel uses the full Type I migration rate from linear theory,
the upper right panel uses 30\% of the full rate, the bottom left panel
uses 10\% of the full rate, and the bottom right panel uses 3\% of the full
rate.  See the electronic edition of the Journal for a color version of this
figure.\label{fig2}}
\end{figure}

\begin{figure}
\plotone{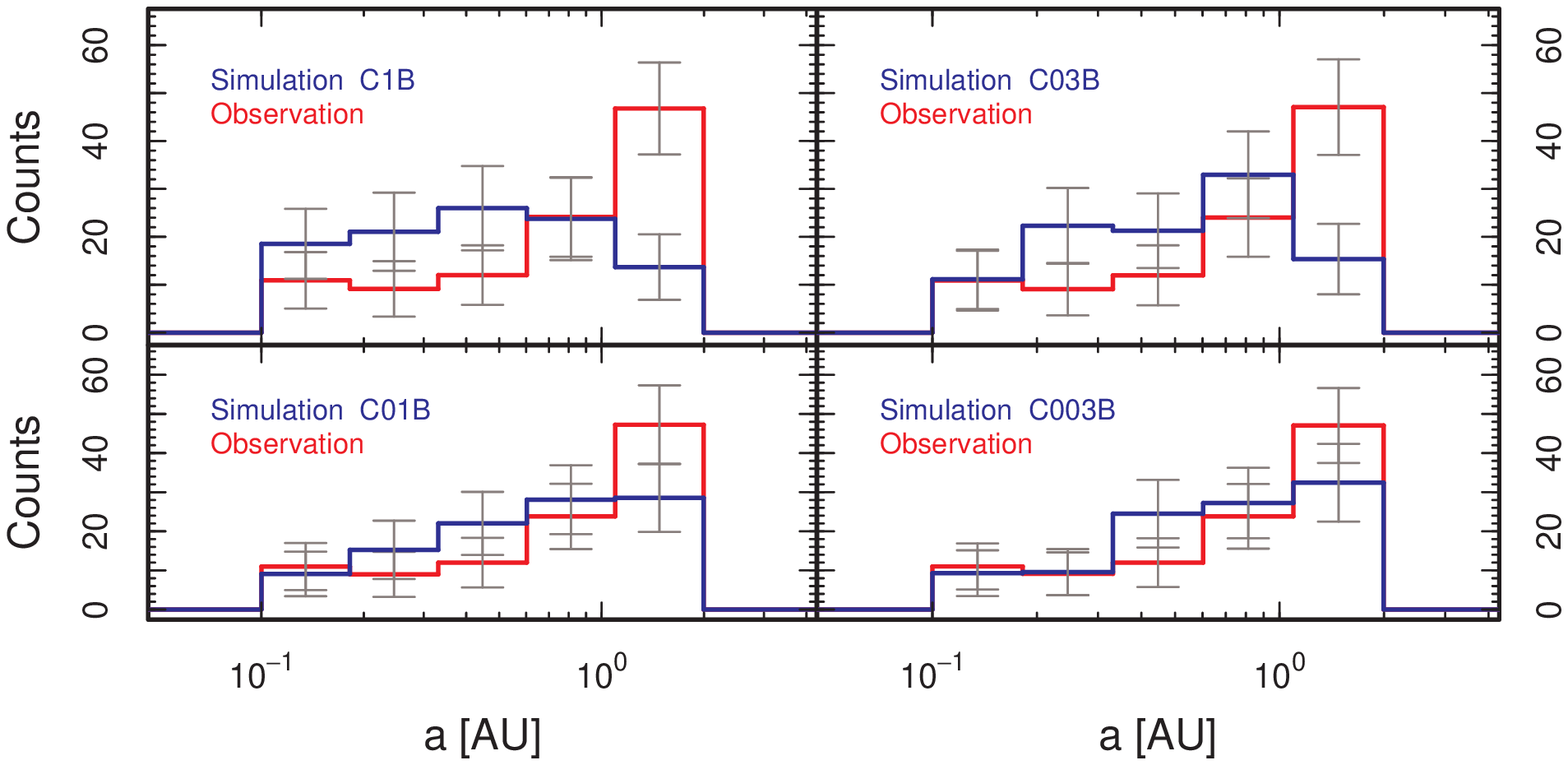}
\caption{1D distribution in semimajor axis derived from the projection
of the 2D distribution in the complete region of the $M_p-a$ plane from
\citet{ida08b} for the disks with the bump in $\Sigma_g$ but without
enhancement in  $\Sigma_d$.
The upper left panel uses the full Type I migration rate from linear theory,
the upper right panel uses 30\% of the full rate, the bottom left panel
uses 10\% of the full rate, and the bottom right panel uses 3\% of the full
rate.  The error bars indicate the 2-$\sigma$ region.  See the electronic
edition of the Journal for a color version of this
figure.\label{fig3}}
\end{figure}

\begin{figure}
\plotone{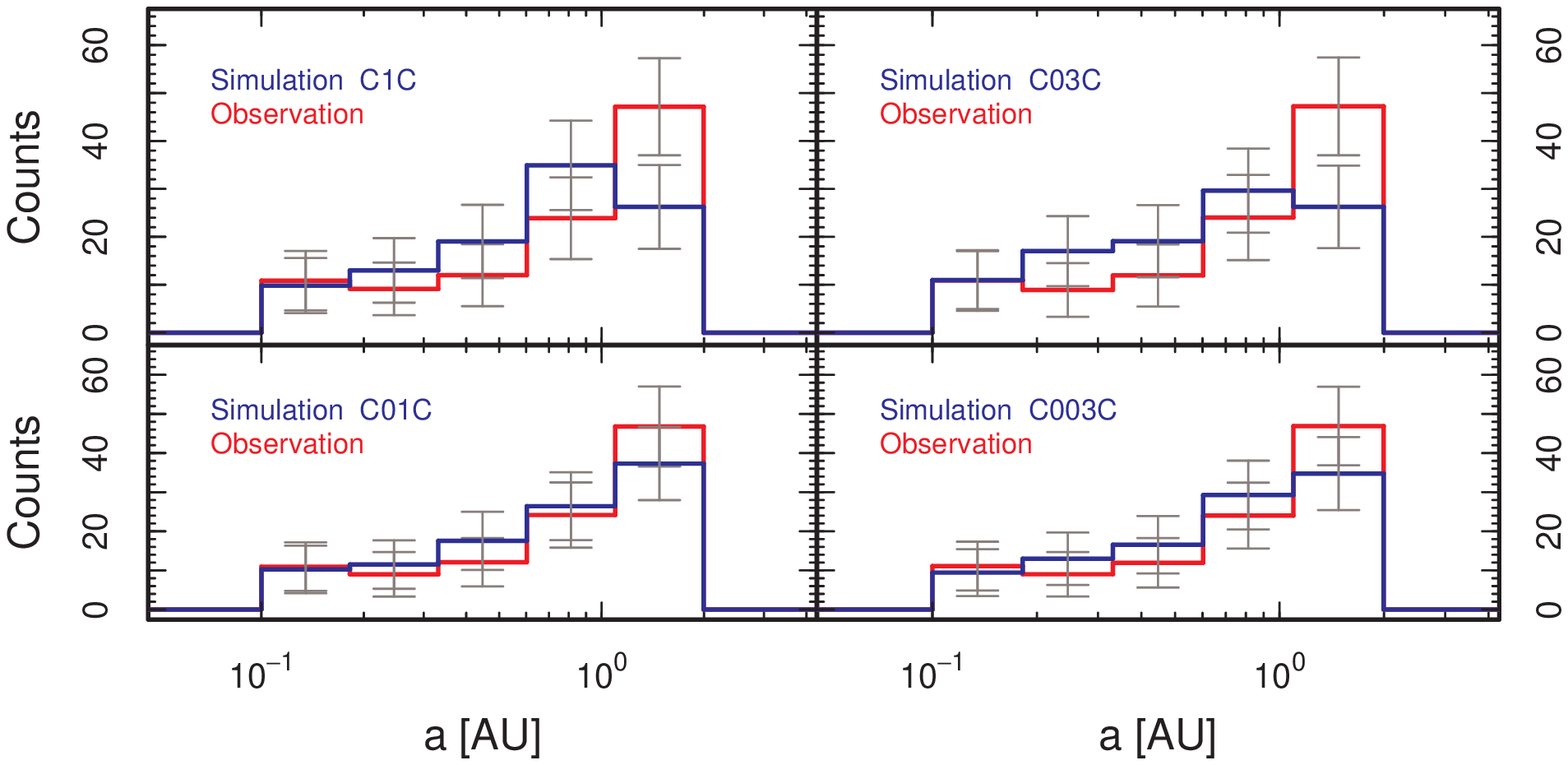}
\caption{1D distribution in semimajor axis derived from the projection
of the 2D distribution in the complete region of the $M_p-a$ plane from
\citet{ida08b} for the disks with the bump in both $\Sigma_g$ and $\Sigma_d$.
The upper left panel uses the full Type I migration rate from linear theory,
the upper right panel uses 30\% of the full rate, the bottom left panel
uses 10\% of the full rate, and the bottom right panel uses 3\% of the full
rate.  The error bars indicate the 2-$\sigma$ region.  See the electronic
edition of the Journal for a color version  of this figure.\label{fig4}}
\end{figure}

\citetalias{ida08b} generated a set of 12 realizations of the $M_p-a$ plane
under different physical assumptions described in Table~\ref{tbl-1}.
One group of models result from disks with the characteristic bump in
gas surface density $\Sigma_g$ due to the coupling effect of the MRI
activity and the ice line and with an enhancement in solid surface
density $\Sigma_d$, another group of models has the bump in $\Sigma_g$
but not in $\Sigma_d$, and the last group ignores the effects the ice
line would have on $\Sigma_g$ and $\Sigma_d$.  For each group of
models, four different parameterizations of the Type I migration rate
were used: 100\%, 30\%, 10\%, and 3\% of the full rate from linear
theory.  For each group of models we marginalize the 2D $M_p-a$
distribution of the simulated exoplanet systems in the complete region
over planetary mass, leaving us with 1D distributions in semimajor
axis.  We plot histograms for each group of models: Figure~\ref{fig2}
corresponds to the models that disregard the effects of the ice line
on $\Sigma_g$ and $\Sigma_d$, Figure~\ref{fig3} corresponds to the
models that includes the effect of the ice line on $\Sigma_g$ but not
$\Sigma_d$, and Figure~\ref{fig4} corresponds to the models that
include the effects of the ice line on $\Sigma_g$ and $\Sigma_d$.

\begin{figure}
\plotone{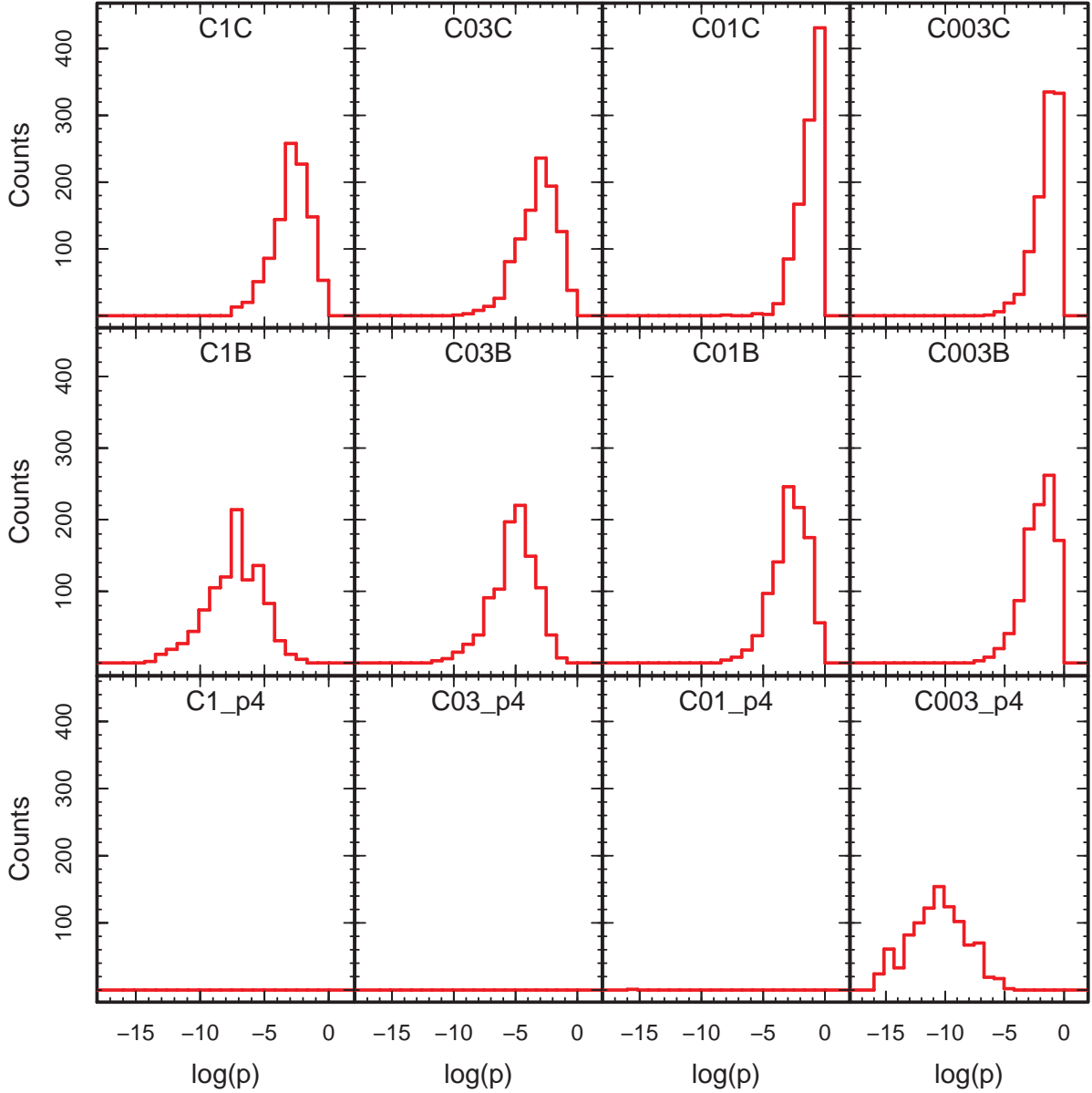}
\caption{Distribution of the Kolmogorov-Smirnov test $p$-values resulting
from 1000 bootstrap resamplings; similar distributions will have a
sharply-peaked $p$-value distributions with a maximum near
$p \sim 1 \Rightarrow \log{p} \sim 0$.  The top row shows the results for
models including the bump in both $\Sigma_g$ and $\Sigma_d$, the middle
row shows models including just the bump in $\Sigma_g$, and the bottom
row shows models with no bump in $\Sigma_g$ or $\Sigma_d$.  For all rows,
the first column shows models with the full Type I migration rate,
the second column shows models with 30\% of the full Type I migration rate,
the third column shows models with 10\% of the full Type I migration rate,
and the fourth column shows models with 3\% of the full Type I migration rate.
The $p$-value distribution of the model with the bump in both
$\Sigma_g$ and $\Sigma_d$ and 10\% of the full Type I migration rate is the
best match to the observed data.  There are no histograms for models C1\_p4,
C03\_p4, or C01\_p4 because the $p$-values were vanishingly small.
See the electronic edition of the Journal for a color version  of this
figure.\label{fig5}}
\end{figure}

We determine the model that best matches the observed $M_p-a$
distribution in the complete region by computing for each model the
$p$-value distribution that results from comparing 1000 bootstrap
resamplings from that model in the complete region with 1000
resamplings of the observed data in the complete region.  We include
the results of this calculation in Figure~\ref{fig5} and we also
report median $p$-values and 95\% intervals in the last three columns
of Table~\ref{tbl-1}.  We find that only models that include the
effects of the ice line on both the gas and solid surface density of the disk
and apply a Type I migration rate an order-of-magnitude less than that
predicted by linear theory (models C01C and C003C) are consistent with the
observed data in the complete region.  Models which neglect the presence of
the ice-line barrier generally do not yield the observed up-turn in the period
distribution of the known planets and are rejected at very high confidence.
Models with efficient Type I migration generally under predict the fraction of
stars with detectable gas giants, especially for those with $a$ outside the
ice line.  Therefore, we argue that the population synthesis models
presented in \citetalias{ida08b} incorporating a Type I migration rate much
reduced from linear theory and the effects of the ice line on both the solid
and gas surface densities ($\Sigma_d$ and $\Sigma_g$ respectively) are at
least plausible deterministic models for giant planet formation.  We also note
that the observed data suggests that the ad hoc prescription for the
location of the ice line used by \citetalias{ida08b} underestimates its radius
by perhaps even a factor of two.

\begin{figure}
\plotone{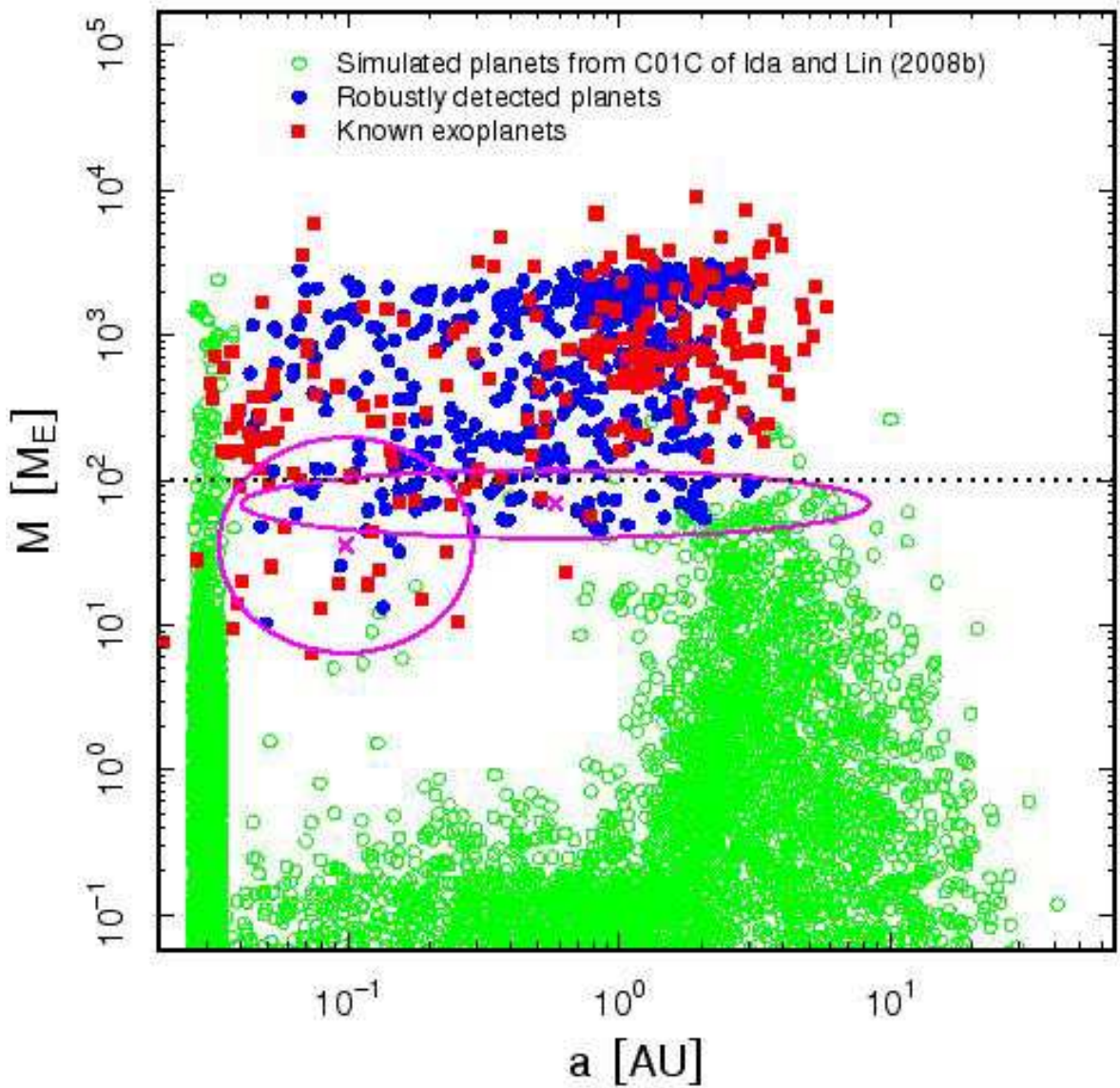}
\caption{Results of our Monte Carlo simulation.  The open circles are
the simulated planets of model C01C.  As in Figure~\ref{fig1}, the filled
circles are planets that are robustly detected by a radial velocity survey
with $n \sim 700$ observations per program star; we plot all known exoplanet
planets as filled squares.  The centers of the best fit Gaussian mixture to
all robustly detected exoplanets below the dashed line at $100~M_{\oplus}$
models are marked by the two X's, while the characteristic ellipses
of each component are the solid lines.  We say
the ``desert" is detected if the mean vectors of the two components
are offset by more than $0.6$ in $\log{a}$ and the minor axis of the 
ellipse at small orbital radius is larger than the minor axis of the
ellipse at large orbital radius.  We find that the ``desert" is detected more
than 90\% of the time when $\mu_n = 700$.  See the electronic edition of the
Journal for a color version  of this figure.\label{fig6}}
\end{figure}

Furthermore, in the presentation of their population synthesis models
\citetalias{ida08b} illustrated the presence of a ``desert" in the $M_p-a$
distribution.  This sparsely populated region is depleted due to both Type I
migration and runaway gas accretion.  For Models C01C and C003C, this
``desert" is centered around $M_p \sim 30~M_{\oplus}$ and $a <1$ AU.
We use the same detection strategy described above, only now we
model the number of radial velocity observations of each planetary
system over a period of about ten years as a Gaussian random variable
with mean $\mu_n$ and standard deviation $\sigma_n$; we round each
random deviate to the nearest whole number.  We then fit a two
component Gaussian mixture model to the $M_p-a$ distribution of all
robustly detected planet with mass $M_p < 100~M_{\oplus}$, and we say
the ``desert" is detected if the mean vectors of the two Gaussians
are offset by more than $0.6$ in $\log{a}$ and the minor axis of the
Gaussian at smaller orbital radius is larger than the minor axis of the
Gaussian at larger orbital radius.  In other words, if the two components of
the mixture model bracket the corner of a region devoid of extrasolar planets
-- a metaphorical ``desert" -- we say the ``desert" is resolved.  We find that
when $\mu_n = 700$ and $\sigma_n = 50$, the two components of the mixture model
bracket a barren region and therefore the ``desert" is resolved more than 90\%
of the time.  We illustrate results of our calculation in Figure~\ref{fig6}.
As a result, a radial velocity campaign with the parameters described above
will be able to confirm the fidelity of the models presented in
\citetalias{ida08b}.

\section{Discussion}

The key prediction of \citetalias{ida08b} is that for masses at which the
dominant migration mechanism is Type I migration -- $M_p \lesssim
50~M_{\oplus}$ -- there will be a dearth of exoplanets within 1 AU of
their host stars, simply because the timescale for Type I migration
is so much shorter than the disk dispersal time. Since they are formed
interior to the ice line, these planets are likely to be rocky and
have mass a few $M_\oplus$. \citetalias{ida08b} also predict an
overdensity of gas giant planets at $\sim 2$ AU resulting from the ice
line.  We show in Figure~\ref{fig6} that one can
quantitatively detect these features in the observed $M_p-a$ with a
radial velocity survey with 1 m s$^{-1}$ precision and about 700
radial velocity observations, or about ten years worth of data from the
Automated Planet Finder (APF).  In future work, we will examine the
ability of missions like the Space Interferometry Mission (SIM) to
verify the same features in $M_p-a$ plane.

In addition to the upper and lower $M_p$ bound, the ``desert" is also
surrounded by populated domains in the $a$ distribution. While the ice
line provides a strong up-turn at a few AU, the models by
\citetalias{ida08b} also imply a large population of short-period rocky
planets as a consequence of Type I migration.  Despite an order-of-magnitude
decrement in the efficiency of Type I migration, the simulated
$M_p-a$ distribution of models C01C and C003C indicate that in the
proximity of their host stars, rocky planets with $M_p \sim$ a few
$M_{\oplus}$ are at least an order-of-magnitude more common than close-in gas
giants (see Figure 6 of \citetalias{ida08b}).  Other authors have already
pointed out the observational difficulties inherent in the search for this
population of ``super-Earths" through radial velocity observations
\citep{nar05}.  Finally, we note that the models of IL did not include
dynamical interactions between planets in multiple planetary systems, and
these interactions can broaden the simulated $a$ distribution and eccentricity
distribution.  We will include these effects in future generations of the
population synthesis models.

\section{Conclusion}

We used the fact that existing exoplanet radial velocity surveys are
complete in the planetary mass-semimajor axis ($M_p-a$) plane where
0.1 AU $< a <$ 2.0 AU and $M_p$ is in the range specified by
Equation~(\ref{eq2}) to show that the observed semimajor axis distribution in
the complete region cannot be explained by models of planet formation that use
the full Type I migration rate predicted by linear theory and that do not
include the effects of the ice line.  Moreover, we also demonstrated that the
expected ``desert" in the $M_p-a$ plane at about $M_p \sim 30~M_{\oplus}$ and
$a <1$ AU predicted by \citetalias{ida08b} will be discernible by a radial
velocity survey with 1 m s$^{-1}$ precision and $n \sim 700$ radial velocity
observations of program stars.  Such an observational campaign will also verify
the predicted inner boundary of the ``desert" where we expect a large
population of super-Earths have migrated to and halted in the proximity of
their host stars.

\acknowledgments We thank A. Cumming, G. Laughlin, G. Marcy, and
Michel Mayor for useful conversation and the anonymous referee for some
insightful comments.  This research has made use of NASA's Astrophysics Data
System Bibliographic Services.  This material is based upon
work supported under a National Science Foundation Graduate Research
Fellowship, NASA (NAGS5-11779, NNG06-GF45G, NNX07A-L13G, NNX07AI88G),
JPL (1270927), NSF(AST-0507424), and JSPS.


\clearpage
\begin{deluxetable}{lrccccc}
\tablecaption{Model Descriptions from \citet{ida08b}\label{tbl-1}}
\tablewidth{0pt}
\tablehead{
\colhead{Name} & \colhead{$C_1$\tablenotemark{a}} &
\colhead{$\Sigma_g$ Enhanced} & \colhead{$\Sigma_d$ Enhanced} &
\colhead{$p_{L}$\tablenotemark{b}} & \colhead{$\bar{p}$\tablenotemark{c}} &
\colhead{$p_{U}$\tablenotemark{d}}}
\startdata
C1C      &    1 & Yes & Yes & $6.9 \times 10^{-6}$ & $2.8 \times 10^{-3}$ &
$1.7 \times 10^{-1}$\\
C03C     &  0.3 & Yes & Yes & $1.6 \times 10^{-6}$ & $1.7 \times 10^{-3}$ &
$1.2 \times 10^{-1}$\\
C01C     &  0.1 & Yes & Yes & $9.9 \times 10^{-3}$ & $8.6 \times 10^{-2}$ &
$7.2 \times 10^{-1}$\\
C003C    & 0.03 & Yes & Yes & $3.2 \times 10^{-3}$ & $6.0 \times 10^{-2}$ &
$6.0 \times 10^{-1}$\\
C1B      &    1 & Yes &  No & $2.9 \times 10^{-12}$ & $7.3 \times 10^{-8}$ &
$9.6 \times 10^{-5}$\\
C03B     &  0.3 & Yes &  No & $5.8 \times 10^{-9}$ & $1.4 \times 10^{-5}$ &
$2.8 \times 10^{-3}$\\
C01B     &  0.1 & Yes &  No & $3.4 \times 10^{-6}$ & $2.8 \times 10^{-3}$ &
$2.2 \times 10^{-1}$\\
C003B    & 0.03 & Yes &  No & $5.1 \times 10^{-5}$ & $1.2 \times 10^{-2}$ &
$3.8 \times 10^{-1}$\\
C1\_p4   &    1 &  No &  No & $1.0 \times 10^{-16}$ & $1.0 \times 10^{-16}$ &
$1.0 \times 10^{-16}$\\
C03\_p4  &  0.3 &  No &  No & $1.0 \times 10^{-16}$ & $1.0 \times 10^{-16}$ &
$1.0 \times 10^{-16}$\\
C01\_p4  &  0.1 &  No &  No & $1.0 \times 10^{-16}$ & $1.0 \times 10^{-16}$ &
$1.0 \times 10^{-16}$\\
C003\_p4 & 0.03 &  No &  No & $4.4 \times 10^{-16}$ & $2.2 \times 10^{-11}$ &
$3.6 \times 10^{-7}$\\
\enddata
\tablenotetext{a}{From \citet{ida08b} -- $C_1$ equivalent to the fraction of
the full Type I migration rate predicted from linear theory applied during
the simulation}
\tablenotetext{b}{Lower bound of an interval centered on median $p$-value
which contains 95\% of our bootstrap realizations}
\tablenotetext{c}{Median $p$-value of our bootstrap realizations}
\tablenotetext{d}{Upper bound of an interval centered on median $p$-value
which contains 95\% of our bootstrap realizations}
\end{deluxetable}

\clearpage


\clearpage


\clearpage


\clearpage


\clearpage


\clearpage


\end{document}